\documentclass[reprint,amsmath,amssymb,aip,jcp]{revtex4-1}

\usepackage{graphicx}
\usepackage{bm}

\usepackage{amsbsy}
\usepackage{color}

    \newcommand{\beq}{\begin{equation}}
    \newcommand{\eeq}{\end{equation}}
    \newcommand\beqa{\begin{eqnarray}}
    \newcommand\eeqa{\end{eqnarray}}
    \newcommand{\nn}{\nonumber\\}
    \newcommand{\la}{\lambda}
    \newcommand{\e}{\eta}
    
    \newcommand{\one}{{(1)}}
    \newcommand{\two}{{(2)}}
    \newcommand{\three}{{(3)}}
    \newcommand{\four}{{(4)}}

\begin{document}



\title{Structural properties of fluids interacting via piece-wise constant potentials with a hard core}



\author{Andr\'es Santos}
\email{andres@unex.es}
\homepage{http://www.unex.es/eweb/fisteor/andres/}

\author{Santos B. Yuste}
\email{santos@unex.es}
\homepage{http://www.unex.es/eweb/fisteor/santos/}

\affiliation{Departamento de F\'{\i}sica, Universidad de
Extremadura, Badajoz, E-06071, Spain}

\author{Mariano L\'opez de Haro}
\email{malopez@unam.mx}
\homepage{http://xml.cie.unam.mx/xml/tc/ft/mlh/}
\affiliation{Instituto
de Energ\'{\i}as Renovables, Universidad Nacional Aut\'onoma
de M\'exico (U.N.A.M.), Temixco, Morelos 62580, M{e}xico}

\author{Mariana B\'arcenas}
\email{mbarcenas@tese.edu.mx}

\affiliation{Divisi\'on de Ingenier\'{\i}a Qu\'{\i}mica y Bioqu\'{\i}mica, Tecnol\'ogico de Estudios Superiores de Ecatepec, Ecatepec, Edo. de M\'exico, 55210, M{e}xico}

\author{Pedro Orea}
\email{porea@imp.mx}

\affiliation{Programa de Ingenier\'{\i}a Molecular, Instituto Mexicano del Petr\'oleo,
M\'exico, D.F., 07730, M{e}xico}

\begin{abstract}
The structural properties of fluids whose molecules interact via potentials with a hard core plus two piece-wise constant sections of different widths and heights are presented. These follow from the more general development previously introduced for potentials with a hard core plus $n$ piece-wise constant sections [Condens. Matter Phys. {\bf 15}, 23602 (2012)] in which use was made of a semi-analytic rational-function approximation method. The results of illustrative cases comprising eight different combinations of wells and shoulders are compared both with simulation data and with those that follow from the numerical solution of the Percus--Yevick and hypernetted-chain  integral equations.
{It is found that  the rational-function approximation generally predicts a more accurate radial distribution function than the Percus--Yevick theory  and is comparable  or even superior to the hypernetted-chain theory. This superiority over both integral equation theories is lost, however, at high densities, especially as the widths of the wells and/or the barriers increase.}

\end{abstract}

\date{\today}


\maketitle

\section{Introduction}
\label{intro}

Due to their relative simplicity, while having at the same time the ability to adequately account for diverse physical features of real fluids, discrete potentials of the form
\begin{equation}
\varphi(r)=\left\{
\begin{array}{ll}
\infty  ,& r<\sigma, \\
\epsilon_1  ,& \sigma<r<\lambda_1 \sigma,\\
\epsilon_2  ,& \lambda_1\sigma<r<\lambda_2 \sigma, \\
\vdots&\vdots \\
\epsilon_n  ,& \lambda_{n-1}\sigma<r<\lambda_n \sigma, \\
0,&r>\lambda_n \sigma,
\end{array}
\right.
\label{varphi}
\end{equation}
have received some attention in the recent literature.\cite{CMA84,CS84,CSD89,BG99,VBG01,FMSBS01,SBFMS04,MFSBS05,BPGS06,CBR07,BFNB08,RPPS08,BNB09,RPSP10,FTR11,GCC07,%
BFGMSSSS02,FMSBS02,FRT06,ZLJ08,VF11,HTS11,BOO11,SYH12,LC12,BOO13} They comprise a hard core of diameter $\sigma$ and $n$ steps of ``heights'' $\epsilon_j$ and widths $(\la_j-\la_{j-1})\sigma$, with $\la_0=1$, so that $\la_n\sigma$ denotes the total range of $\varphi(r)$. The sign of $\epsilon_j$ determines whether the $j$th step is either a ``shoulder'' ($\epsilon_j>0$) or a ``well'' ($\epsilon_j<0$). The interaction potential at $r=\la_j\sigma$ ($j=1, 2,\ldots, n$) is repulsive if $\epsilon_j>\epsilon_{j+1}$ and attractive if $\epsilon_j<\epsilon_{j+1}$ (with the convention $\epsilon_{n+1}=0$). Particular cases of these discrete potentials when $n=1$ are the popular square-well and square-shoulder  potentials.

The phase diagram and the thermodynamic properties of discrete-potential fluids have been thoroughly examined and are relatively well understood.\cite{GCC07,BG99,VBG01,FMSBS01,SBFMS04,MFSBS05,BPGS06,CBR07,BFNB08,RPPS08,BNB09,RPSP10,FTR11}
{On the other hand, although many studies of their structural properties either theoretical or from simulation have been also performed,\cite{BFGMSSSS02,FMSBS02,FRT06,GCC07,ZLJ08,VF11,HTS11,BOO11,SYH12,LC12,BOO13} the variety of cases that may present justifies further work on this subject.}

In a previous paper,\cite{SYH12} following a {semi-analytic} methodology referred to generically as the rational-function approximation (RFA) that, although approximate, has proven successful for many other systems,\cite{HYS08} we derived the general formulae for the structural properties of fluids whose molecules interact via discrete potentials with a hard core plus an arbitrary number of piece-wise constant sections of different widths and heights. The theoretical scheme was illustrated by comparing it with available computer simulations results.\cite{BOO11}

The aim of this paper is to carry out a more systematic study of the structural properties of fluids characterized by a discrete potential with a hard core plus different combinations of a repulsive shoulder and an attractive well. This will be done by considering the results of Ref.\ \onlinecite{SYH12} in the case of $n=2$ for various values of the parameters and subsequently performing a comparison both with simulation results as well as with those  that follow from the numerical solution of the Ornstein--Zernike (OZ) equation with both the Percus--Yevick (PY) and hypernetted-chain (HNC) closures.
{As will be seen, the performance of the RFA approach, despite its simplicity, is quite satisfactory.}

The paper is organized as follows. In order to make it self-contained, in Sec.\ \ref{sec2} we introduce the systems to be studied and sketch the derivation of the results of Ref.\ \onlinecite{SYH12} for the structural properties of such systems when $n=2$. We also include here some details of the simulation and of the numerical solution of the OZ equation with the PY and HNC closures. This is followed in Sec.\ \ref{sec3} by the comparison between the outcomes of the three different approaches for the radial distribution function (RDF). The paper is closed in Sec.\ \ref{sec4} with further discussion and some concluding remarks.

\section{System and structural properties}
\label{sec2}

We consider a fluid of number density $\rho$ and absolute temperature $T$ in which the intermolecular pair potential is of the form of Eq.\ \eqref{varphi} with $n=2$. We will take the hard-core diameter $\sigma$ as the length unit so all distances will be measured in units of $\sigma$.
The two {main  quantities usually employed to characterize the structure of fluids in equilibrium} are the static structure factor $S(q)$ and the RDF $g(r)$ which are related by
\begin{eqnarray}
\label{b1}
S(q)&=&1+\rho \int {d} {r}\, {e}^{-{i} \mathbf{q}\cdot \mathbf{r}} [g(r)-1]\nonumber\\
&=&1-2\pi\rho \left.\frac{G(s)-G(-s)}{s}\right|_{s={i} q},
\end{eqnarray}
where
\begin{equation}
\label{b3}
G(s)=\int_0^\infty {d} r\, {e}^{-rs} rg(r)
\end{equation}
is the Laplace transform of $rg(r)$.

\subsection{The rational-function approximation method}
We define an
auxiliary function $F(s)$ directly related to $G(s)$ through
\begin{eqnarray}
\label{b5}
G(s)&=&s\frac{F(s)e^{-s}}{1+12\eta F(s)e^{-s}}\nonumber\\
 &=&\sum_{m=1}^\infty (-12\eta)^{m-1}s[F(s)]^m e^{-ms}.
\end{eqnarray}
Here, $\eta=(\pi/6)\rho\sigma^3$ is the packing fraction. Laplace inversion of Eq.\ (\ref{b5}) provides a useful representation of
$g(r)$, namely
 \begin{equation}
 \label{b6}
 g(r)=r^{-1}\sum_{m=1}^\infty (-12\eta)^{m-1}f_m(r-m)\Theta(r-m),
 \end{equation}
where $f_m(r)$ is the inverse Laplace transform of $s[F(s)]^m$ and $\Theta(r)$
is the Heaviside step function.

The contact value $g(1^+)$ of the RDF  is related to $F(s)$ through $g(1^+)=f_1(0)=\text{lim}_{s\rightarrow \infty} s^2F(s)$ and it has to be finite. Further, {as seen from Eq.\ \eqref{b1},} the behavior of $G(s)$ for small $s$ determines the value of $S(0)$, which must also be finite. Hence, $F(s)$ must satisfy two conditions:\cite{SYH12}
\begin{equation}
\label{b9}
F(s)\sim s^{-2},\quad s\rightarrow \infty,
\end{equation}
\beqa
\label{b10}
F(s)&=&-\frac{1}{12\eta}\left(1+s+\frac{1}{2}s^2+\frac{1+2\eta}{12\eta}s^3
+\frac{2+\eta}{24\eta}s^4\right)\nn
&&+{\cal O}(s^5).
\eeqa

Equations \eqref{b5}--\eqref{b10} are exact and valid for any interaction potential with a hard core at $r=\sigma=1$. Now we particularize to the potential \eqref{varphi} with $n=2$.
To reflect the discontinuities of $g(r)$ at the points $r=\la_1$ and $r=\la_2$, where $\varphi(r)$ is discontinuous, we decompose $F(s)$ as
\beq
F(s)=R_0(s)+R_1(s)e^{-(\lambda_1-1)s}+R_2(s)e^{-(\lambda_2-1)s}.
\label{3.1}
\eeq
As a consequence,
\beqa
f_1(r)&=&\xi_0(r)\Theta(r)+\xi_1(r-\la_1+1)\Theta(r-\la_1+1)\nn
&&+\xi_2(r-\la_2+1)\Theta(r-\la_2+1),
\label{xi}
\eeqa
where $\xi_j(r)$ denotes the inverse Laplace transform of $sR_j(s)$.
If, as will be done here, one assumes that $\lambda_2\leq 2$, insertion of Eq.\ \eqref{xi} into Eq.\ \eqref{b6} gives the RDF in the shell $1<r<2$. In particular,
\beq
g(\la_1^-)=\la_1^{-1}\xi_0(\la_1-1),
\label{la1m}
\eeq
\beq
g(\la_1^+)=\la_1^{-1}\left[\xi_0(\la_1-1)+\xi_1(0)\right],
\label{la1p}
\eeq
\beq
g(\la_2^-)=\la_2^{-1}\left[\xi_0(\la_2-1)+\xi_1(\la_2-\la_1)\right],
\label{la2m}
\eeq
\beq
g(\la_2^+)=\la_2^{-1}\left[\xi_0(\la_2-1)+\xi_1(\la_2-\la_1)+\xi_2(0)\right].
\label{la2p}
\eeq

Now we assume the following \emph{rational-function} approximation for $R_j(s)$:
\beq
R_j(s)=-\frac{1}{12\eta}\frac{A_j+B_j s}{1+S_1 s+S_2 s^2+S_3 s^3}, \quad j=0,1,2.
\label{c6}
\eeq
Note that Eq.\ \eqref{c6} for $R_0(s)$ guarantees the fulfillment of the physical condition \eqref{b9}. Yet, the approximation \eqref{c6} contains nine parameters to be determined.
The exact expansion \eqref{b10} imposes five constraints among those nine parameters, namely\cite{SYH12}
\beq
1=A_0+ A_1+A_2,
\label{c7}
\eeq
\beq
S_1=-1+B_0-C^\one,
\label{c8}
\eeq
\beq
S_2=\frac{1}{2}-B_0+C^\one+\frac{1}{2}C^\two,
\label{c9}
\eeq
\beq
S_3=-\frac{1+2\e}{12\e}+\frac{1}{2}B_0-\frac{1}{2}C^\one-\frac{1}{2}C^\two-\frac{1}{6}C^\three,
\label{c10}
\eeq
\beq
B_0=C^\one+\frac{\e/2}{1+2\e}\left(6C^\two+4C^\three+C^\four\right)+\frac{1+\e/2}{1+2\e}.
\label{c11}
\eeq
Here,
\beq
C^{(k)}\equiv \sum_{j=1}^2 \left[A_j (\lambda_j-1)^k -k B_j(\lambda_j-1)^{k-1}\right].
\label{c12}
\eeq
Next, since the cavity function $y(r)\equiv g(r)e^{\beta \varphi(r)}$, where $\beta\equiv {1}/{k_B T}$ ($k_B$ being the Boltzmann constant), must be continuous at $r=\la_1$ and $r=\la_2$, one obtains from Eqs.\ \eqref{la1m}--\eqref{la2p} the two conditions\cite{SYH12}
\beqa
\frac{B_1}{S_3}&=&\left[e^{\beta(\epsilon_{1}-\epsilon_2)}-1\right]\sum_{\nu=1}^3\frac{s_\nu e^{(\la_1-1) s_\nu}}{S_1 +2S_2 s_\nu+3S_3 s_\nu^2}\nn
&&\times (A_0+B_0s_\nu),
\label{3.5}
\eeqa
\beqa
\frac{B_2}{S_3}&=&\left(e^{\beta\epsilon_{2}}-1\right)\sum_{\nu=1}^3\frac{s_\nu e^{(\la_2 -1)s_\nu}}{S_1 +2S_2 s_\nu+3S_3 s_\nu^2}\nn
&&\times\left[A_0+B_0s_\nu+(A_1+B_1s_\nu)e^{-(\la_1-1) s_\nu}\right],
\label{3.6}
\eeqa
where $s_\nu$ ($\nu=1,2,3$) are the three roots of the cubic equation
\beq
1+S_1 s_\nu+S_2 s_\nu^2+S_3 s_\nu^3=0.
\eeq

Equations \eqref{c7}--\eqref{c11}, \eqref{3.5}, and \eqref{3.6} still leave two parameters undetermined.
A simplifying assumption is that the coefficients $A_j$  ($j=0,1, 2$) may be fixed at their zero-density values, namely
\beq
A_0=e^{-\beta\epsilon_1},\quad A_1=e^{-\beta\epsilon_2}-e^{-\beta\epsilon_1},\quad A_2=1-e^{-\beta\epsilon_2}.
\label{3.2}
\eeq
This closes the problem of determining the nine parameters in terms of $\eta$, $\la_1$, $\la_2$, $\beta \epsilon_1$, and $\beta \epsilon_2$.  In fact, Eqs.\ \eqref{c8}--\eqref{c11} allow us to express $S_1$, $S_2$, $S_3$, and $B_0$ as linear combinations of $B_1$ and $B_2$, so that in the end  one only has to solve (numerically) the two coupled transcendental equations \eqref{3.5} and \eqref{3.6}.
{Since the dependence of $G(s)$ on $s$ is explicit, we} are now in a position to compute the structural quantities of our systems. {The structure factor $S(q)$ can be directly obtained from Eq.\ \eqref{b1}, while the RDF $g(r)$ can be obtained from Eq.\ \eqref{b6} or, more directly, by numerical inverse Laplace transform of $G(s)$}.\cite{AW92}

{It is worth remarking that, while the choice \eqref{3.2} guarantees that the RFA is exact in the limit $\rho\to 0$, it differs from the exact result to first order in density, as discussed in Ref.\ \onlinecite{YS94} for the square-well potential.}

\subsection{The PY and HNC approximations}

In the usual integral equation approach to the theory of liquids, the OZ equation, which may be formally considered as a definition of the direct correlation function $c(r)$, provides a
link between this direct correlation function and the total correlation function $h(r)\equiv g(r)-1$, the latter being  a measure of the `influence',  either direct or through a third molecule, of two molecules separated a distance $r$ away. The OZ relation reads
\beqa
h(r)&=&c(r)+\rho\int d\mathbf{r'}c(r')
h(|\mathbf{r}-\mathbf{r'}|)\nn
&=&c(r)+\frac{2\pi\rho}{r}\int_0^\infty dr'\,{r'}{c(r')}\int_{|r-r'|}^{r+r'}dr''\, {r''}{h(r'')},\nn
  \label{OZ}
\eeqa
where in the second equality we have particularized to three-dimensional systems and used {bipolar coordinates.\cite{B61,BNS08}}

Since both $h(r)$ and $c(r)$ are unknown, in order to close the description one requires an additional equation, known as the closure relation. A closure can be expressed as a local relationship between the direct correlation function, the Mayer function $f(r)\equiv e^{-\beta\varphi(r)}-1$, and the cavity function $y(r)$, i.e.,
\beq
c(r)=\mathcal{C}\left(f(r),y(r)\right).
\label{clos1}
\eeq
Equivalently, Eq.\ \eqref{clos1} can be inverted to obtain a local relationship between the cavity function and the indirect correlation function $\gamma(r)\equiv h(r)-c(r)$, i.e.,
\beq
y(r)=\mathcal{Y}\left(\gamma(r)\right).
\label{clos2}
\eeq
Insertion of the closure \eqref{clos1} and \eqref{clos2} into the OZ relation \eqref{OZ} yields a closed nonlinear integral equation for the cavity function:
\beqa
y(r)&=&\mathcal{Y}\left(\frac{2\pi\rho}{r}\int_0^\infty dr'\,{r'}\mathcal{C}\left(f(r'),y(r')\right)\right.\nn
&&\left.\times\int_{|r-r'|}^{r+r'}dr''\, {r''}\left[{e^{-\beta\varphi(r'')}y(r'')-1}\right]\right).
\label{PY&HNC}
\eeqa

As mentioned in Sec.\ \ref{intro}, we will consider here both the PY and HNC closures given by
\begin{equation}
c(r)=h(r)-y(r)+1 \quad (\text{PY}),
\label{PYC}
\end{equation}
\begin{equation}
c(r)= h(r)-\ln y(r) \quad (\text{HNC}).
\label{HNCC}
\end{equation}
In terms of the functions $\mathcal{C}(f,y)$ and $\mathcal{Y}(\gamma)$, the PY and HNC closure relations are
\beq
\mathcal{C}(f,y)=fy,\quad \mathcal{Y}(\gamma)=1+\gamma \quad \text{(PY)},
\eeq
\beq
\mathcal{C}(f,y)=(f+1)y-1-\ln y,\quad \mathcal{Y}(\gamma)=e^\gamma \quad \text{(HNC)}.
\eeq
{Note that the PY closure can be obtained from the HNC one by formally linearizing $\mathcal{C}(f,y)$ and $\mathcal{Y}(y)$ with respect to $\gamma$. In contrast to the RFA, the PY and HNC theories provide the exact RDF  to first order in density.}

We have solved Eq.\ \eqref{PY&HNC} numerically in the PY and HNC cases. First, a  discretization scheme $y(r)\to \{r_i,y_i\}$ with $r_i=i\Delta r$ ($i=1,2,\ldots,\mathcal{N}$) and a cut-off distance $r_\mathcal{N}=\mathcal{N}\Delta r$ are introduced, so that the integral equation \eqref{PY&HNC} is replaced by a  set of $\mathcal{N}$ nonlinear coupled equations:
\beqa
y_i&=&\mathcal{Y}\left(\frac{2\pi\rho}{r_i}(\Delta r)^2\sum_{j=1}^\mathcal{N} {r_j}\mathcal{C}\left(f_j,y_j\right)\right.\nn
&&\left.\times\sum_{k=|i-j|+1}^{\min(i+j,\mathcal{N})}{r_k}\left({e^{-\beta\varphi_k}y_k-1}\right)\right),\quad i=1,2,\ldots,\mathcal{N}.\nn
\label{PY&HNC2}
\eeqa
Next, a coarse-grained solution of Eq.\ \eqref{PY&HNC2} is obtained by an iteration method. Insertion of the $n$th order input $\{y_i^{(n,\text{in})}\}$ into the right-hand side of Eq.\ \eqref{PY&HNC2} gives the $n$th order output $\{y_i^{(n,\text{out})}\}$ and the subsequent input is constructed as $y_i^{(n+1,\text{in})}=\alpha y_i^{(n,\text{out})}+(1-\alpha)y_i^{(n,\text{in})}$, with a convenient choice of the mixing parameter $\alpha$. The iterations are continued until the convergence criterion
\beq
\max_i|y_i^{(n,\text{out})}-y_i^{(n,\text{in})}|<10^{-3}
\label{conv}
\eeq
is reached or the number of iterations exceeds 100. Once the coarse-grained solution has been obtained, a fine-grained solution of the set of $\mathcal{N}$ equations \eqref{PY&HNC2} is determined with the help of a computational software program, using the coarse-grained solution as a seed.\cite{note_13_04_2}
In the numerical solutions presented in Sec.\ \ref{sec3} we have generally used $\Delta r=0.01\sigma$ and $\mathcal{N}=400$. As for the mixing parameter $\alpha$, it was chosen by trial and error, being  necessary  to decrease it as the density increases.\cite{note_13_07}

\subsection{Technical simulation details}

The simulation data were computed with a Replica Exchange Monte Carlo  (REMC) method. The REMC method is also known as Parallel Tempering and  was derived to achieve good sampling of systems that present a free energy landscape with many local minima.\cite{FS02,OJO11} The REMC method consists of simulating $M$ replicas (copies) of the system at different thermodynamic conditions; the attempted MC moves are accepted or rejected according to the traditional Metropolis algorithm.  Due to these exchanges, a particular replica travels through many temperatures, allowing it to overcome any barriers to free energy.

The method samples an expanded canonical ensemble, taking the temperature as the expansion variable. The existence of this expanded ensemble justifies the introduction of movements of exchange between replicas. The expanded ensemble is defined as
\begin{equation}
Q_{\text{expanded}}=\prod_{i=1}^{M} Q_i,
\label{CEE}
\end{equation}
where $Q_{i}$ is the partition function of the (NVT) canonical ensemble of the system (subensemble $i$) at temperature $T$, volume $V$, and number of particles   $N$. To satisfy the detailed balance condition, the probability of acceptance of the exchange is given by
\begin{equation}
P_\text{acc}=\min(1,\exp[-(\beta_j-\beta_i)(U_j-U_i) ] ),
\label{Accept}
\end{equation}
where $\beta_j-\beta_i$ is the difference between the reciprocal temperatures and $U_j-U_i$ is the  difference between the potential energies of the subensembles $i$ and $j$.

A cubic simulation box of dimensions $L_x = L_y = L_z = 10\sigma$ was used  and periodic boundary conditions were set in the three directions. Verlet lists\cite{V67} were implemented to improve performance. We have carried out computer experiments for different systems, corresponding to different values of the parameters of the potentials that will be specified later. These systems are further characterized by their reduced density $\rho^*=\rho\sigma^3$ and their reduced temperature $T^*=k_B T/\epsilon$, where $\epsilon=\text{max}(|\epsilon_1|,|\epsilon_2|)$. The number of replicas {$M=12$} was chosen to match the number of different temperatures in which we want to examine the systems. The highest temperature was set at $T^* = 2$, while the other temperatures were established following a decreasing geometric progression, {namely $T_n^*=2 \times 0.959^{n-1}$, $n=1,\ldots,12$.} The initial configuration of each system, consisting of a collection of $N=500$ particles randomly arranged in the simulation box and thus setting the reduced number density of our systems as $\rho^*=0.5$, was equilibrated by conducting $10^7$ Monte Carlo simulation steps. The RDF was calculated  over additional $4 \times 10^7$ configurations.

\section{Results}
\label{sec3}

\begin{figure}
\includegraphics[width=8cm]{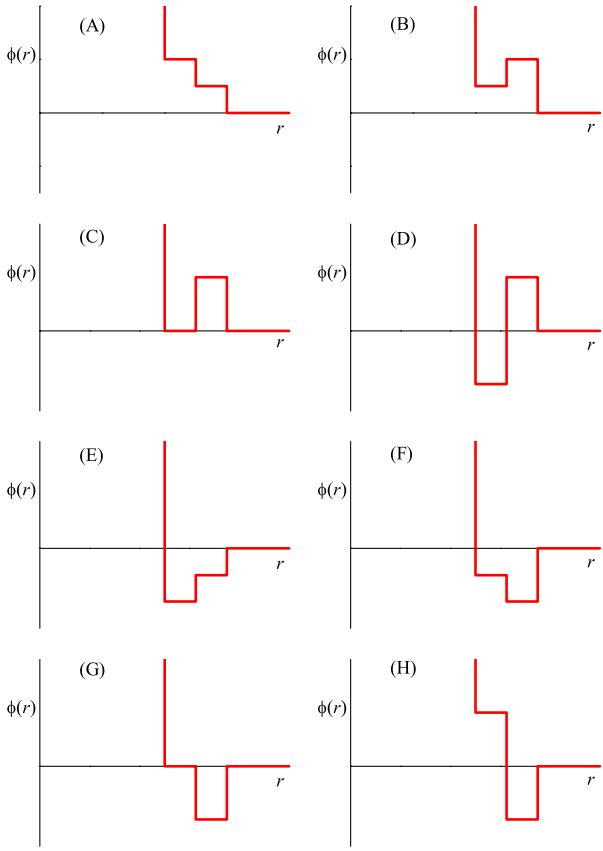}
\caption{ {Diagrams} of the surveyed potentials, labeled from A to H.}
\label{fig2}
\end{figure}

For convenience and in order to try to be systematic, we now first fix the values of $\lambda_1$ and $\lambda_2$ to be $\lambda_1=1.25$ and $\lambda_2=1.5$. {As for the values of $\epsilon_1$ and $\epsilon_2$, we have considered eight representative cases}.
Since, as already stated, $\epsilon=\text{max}(|\epsilon_1|,|\epsilon_2|)$, at least one of the $|\epsilon_i|$ must be equal to $\epsilon$. The other energy level has been chosen as $0$, $\pm \epsilon$, or $\pm \frac{1}{2}\epsilon$ with the following conditions. First, the cases having $\epsilon_1=\epsilon_2$ have not been considered since they correspond to having just one step. The same can be said about the cases with $\epsilon_2=0$.  Finally, if $\epsilon_1$ and $\epsilon_2$ have opposite signs, then we have taken $|\epsilon_1|=|\epsilon_2|=\epsilon$.
The potentials of the different cases (from A to H) are represented graphically in Fig.\ \ref{fig2}.
We observe that system A corresponds to a purely repulsive  potential. In the cases B--D the potential is repulsive {at $r/\sigma=\la_2$ and attractive at $r/\sigma=\la_1$}, with increasing attraction when going from B to D. System E presents a purely attractive tail beyond the hard core. As for systems F--H, the potential is attractive  {at $r/\sigma=\la_2$ and repulsive at $r/\sigma=\la_1$}, with increasing repulsion when going from F to H. {The cases A and F--H are examples of core-softened potentials.}

\begin{figure}
\includegraphics[width=8cm]{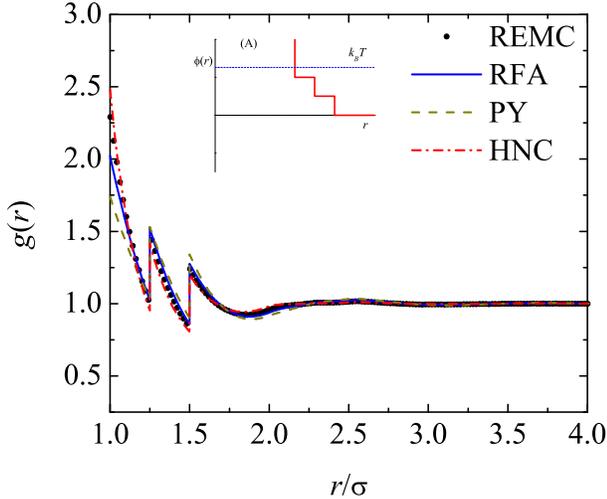}
\caption{ Comparison of the different theoretical approaches to compute the RDF of the system corresponding to case A ($\lambda_1=1.25$, $\la_2=1.5$, $\epsilon_1=\epsilon$, $\epsilon_2=\epsilon/2$) at $\rho^*=0.5$ and $T^*=1.26193$ with simulation results. Solid line: RFA; dashed line: HNC; dotted-dashed line: PY; symbols: REMC data. The inset shows the interaction potential and the temperature value.}
\label{fig3}
\end{figure}

\begin{figure}
\includegraphics[width=8cm]{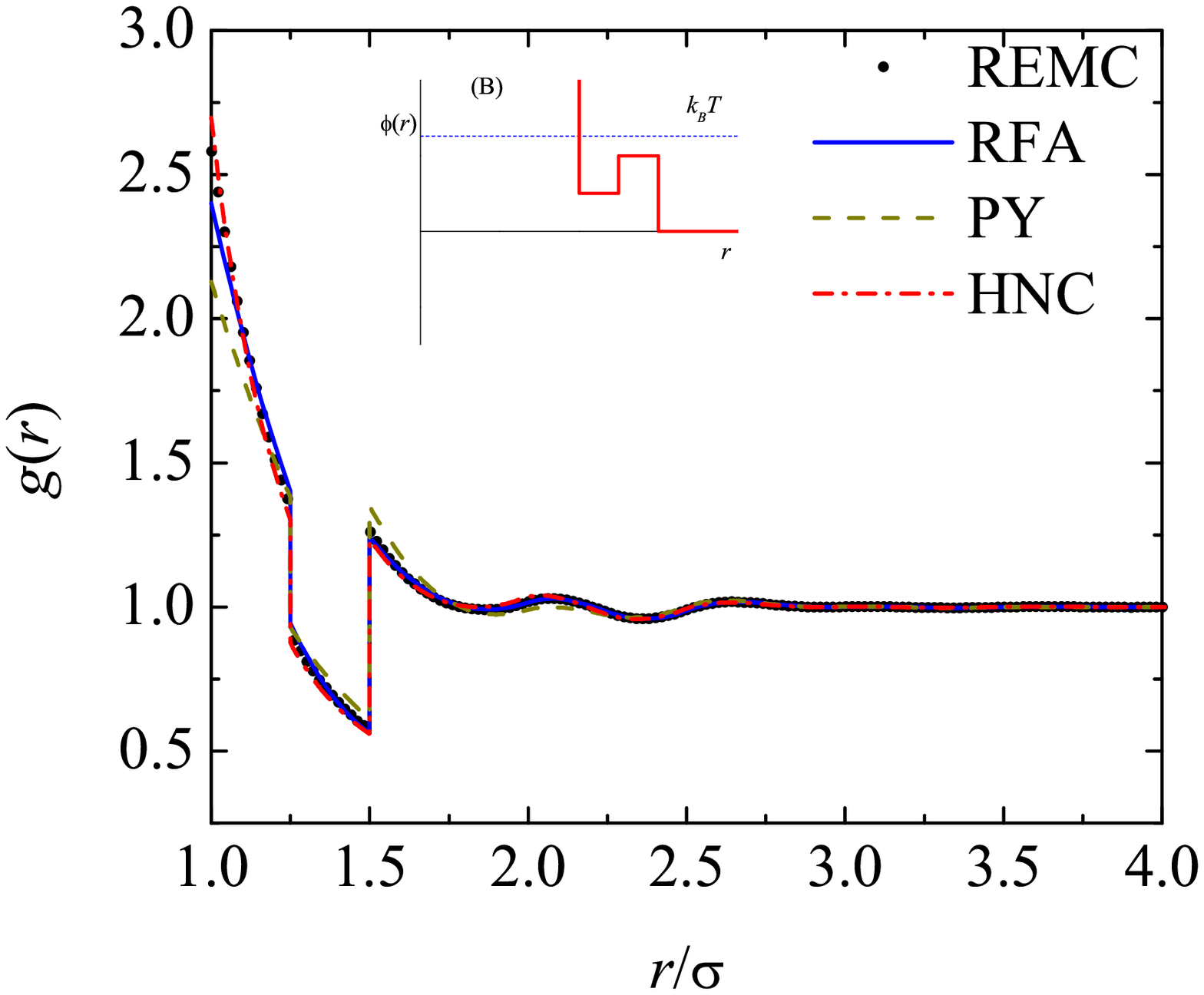}
\caption{ Comparison of the different theoretical approaches to compute the RDF of the system corresponding to case B ($\lambda_1=1.25$, $\la_2=1.5$, $\epsilon_1=\epsilon/2$, $\epsilon_2=\epsilon$) at $\rho^*=0.5$ and $T^*=1.26193$ with simulation results. Solid line: RFA; dashed line: HNC; dotted-dashed line: PY; symbols: REMC data. The inset shows the interaction potential and the temperature value.}
\label{fig4}
\end{figure}

\begin{figure}
\includegraphics[width=8cm]{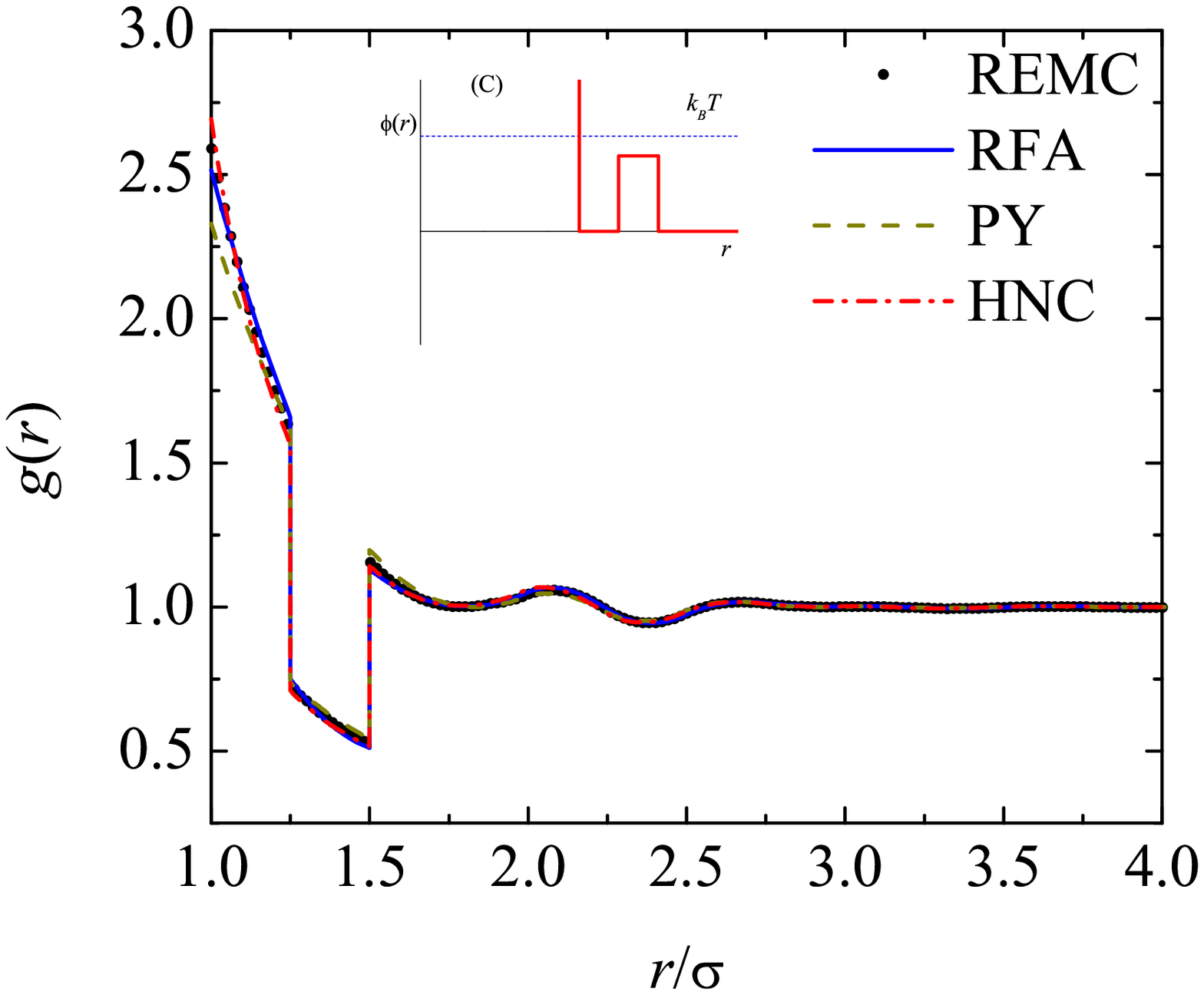}
\caption{ Comparison of the different theoretical approaches to compute the RDF of the system corresponding to case C ($\lambda_1=1.25$, $\la_2=1.5$, $\epsilon_1=0$, $\epsilon_2=\epsilon$) at $\rho^*=0.5$ and $T^*=1.26193$ with simulation results. Solid line: RFA; dashed line: HNC; dotted-dashed line: PY; symbols: REMC data. The inset shows the interaction potential and the temperature value.}
\label{fig5}
\end{figure}

\begin{figure}
\includegraphics[width=8cm]{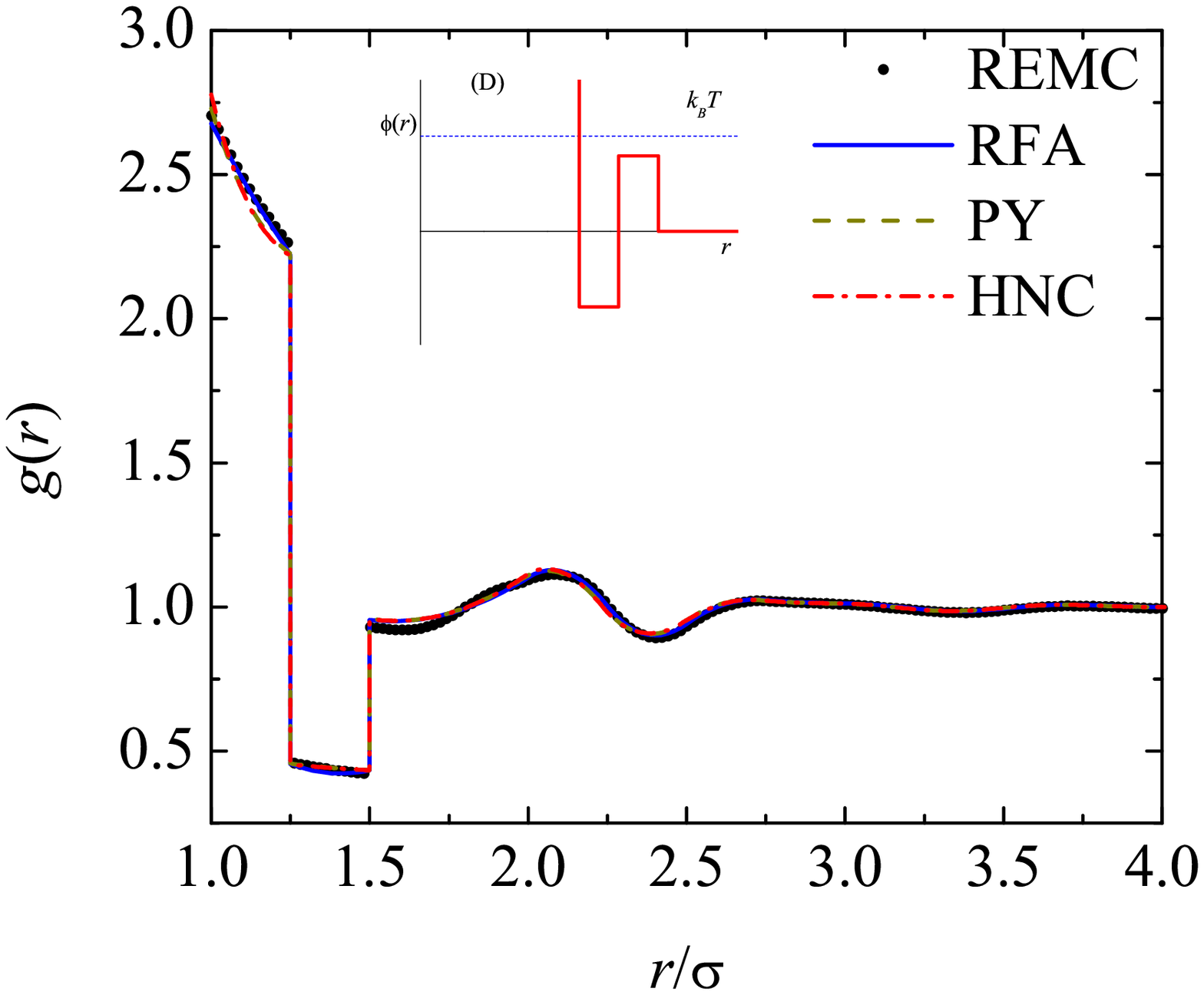}
\caption{ Comparison of the different theoretical approaches to compute the RDF of the system corresponding to case D ($\lambda_1=1.25$, $\la_2=1.5$, $\epsilon_1=-\epsilon$, $\epsilon_2=\epsilon$) at $\rho^*=0.5$ and $T^*=1.26193$ with simulation results. Solid line: RFA; dashed line: HNC; dotted-dashed line: PY; symbols: REMC data. The inset shows the interaction potential and the temperature value.}
\label{fig6}
\end{figure}

\begin{figure}
\includegraphics[width=8cm]{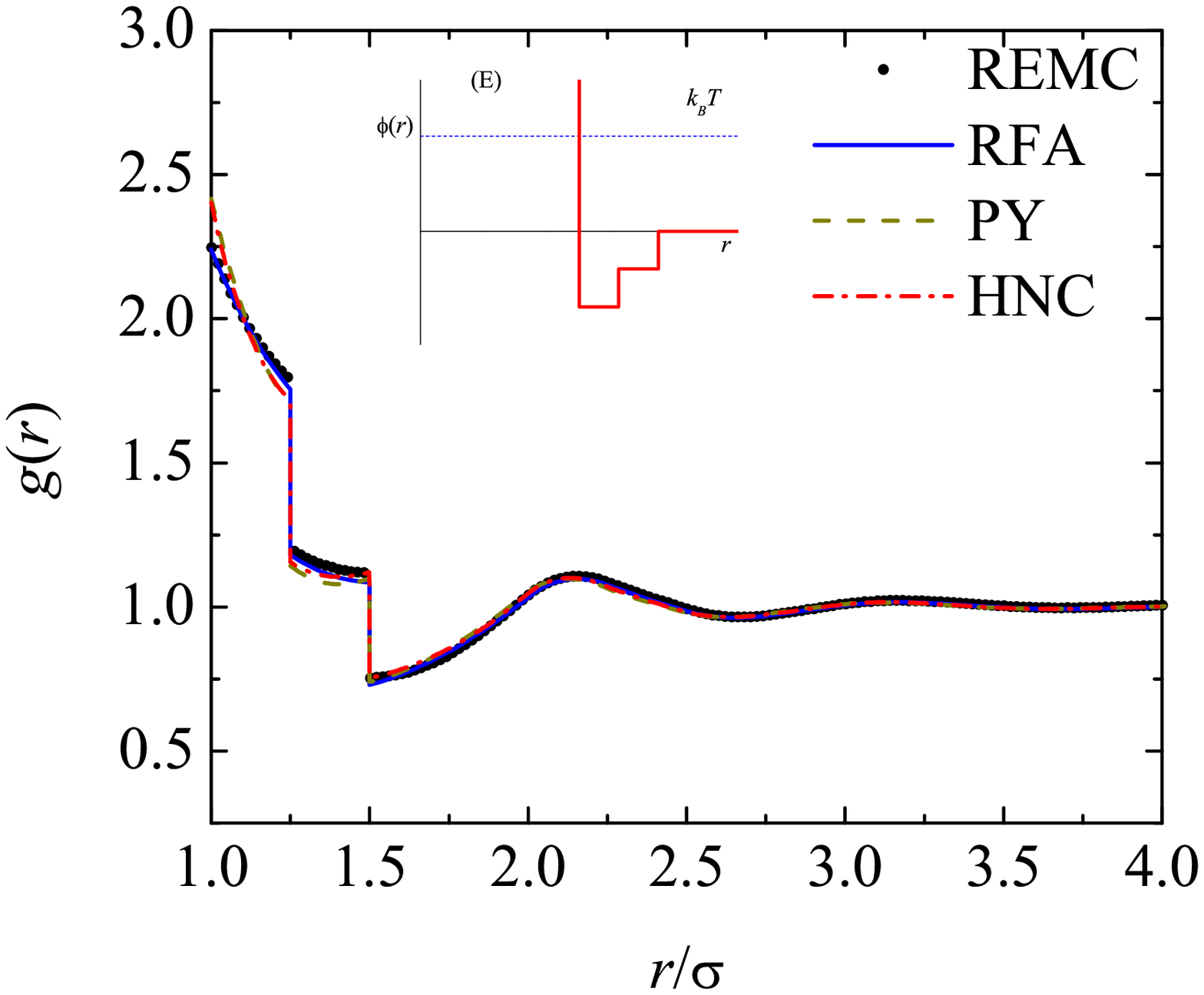}
\caption{ Comparison of the different theoretical approaches to compute the RDF of the system corresponding to case E ($\lambda_1=1.25$, $\la_2=1.5$, $\epsilon_1=-\epsilon$, $\epsilon_2=-\epsilon/2$) at $\rho^*=0.5$ and $T^*=1.26193$ with simulation results. Solid line: RFA; dashed line: HNC; dotted-dashed line: PY; symbols: REMC data. The inset shows the interaction potential and the temperature value.}
\label{fig7}
\end{figure}

\begin{figure}
\includegraphics[width=8cm]{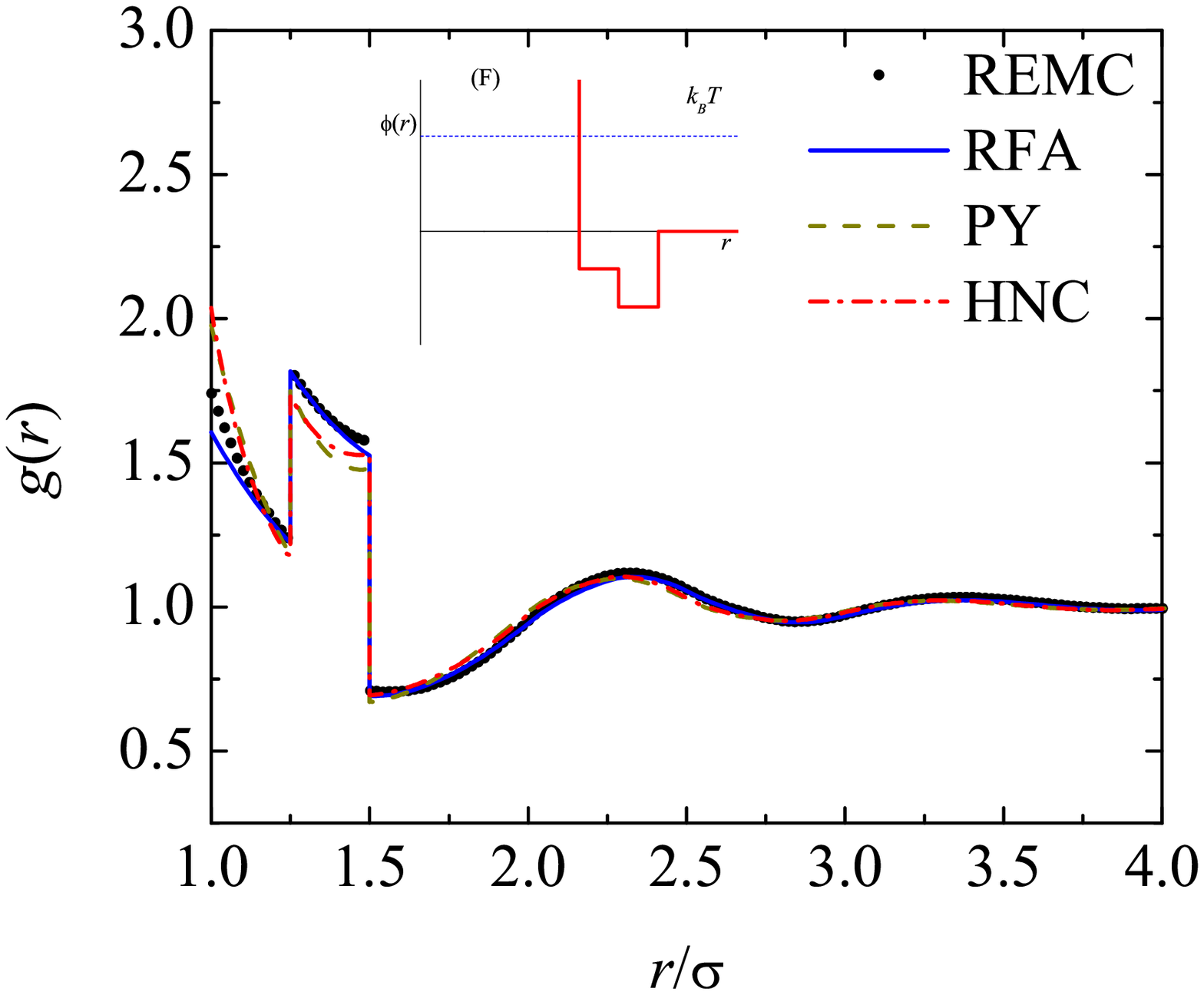}
\caption{ Comparison of the different theoretical approaches to compute the RDF of the system corresponding to case F ($\lambda_1=1.25$, $\la_2=1.5$, $\epsilon_1=-\epsilon/2$, $\epsilon_2=-\epsilon$) at $\rho^*=0.5$ and $T^*=1.26193$ with simulation results. Solid line: RFA; dashed line: HNC; dotted-dashed line: PY; symbols: REMC data. The inset shows the interaction potential and the temperature value.}
\label{fig8}
\end{figure}

\begin{figure}
\includegraphics[width=8cm]{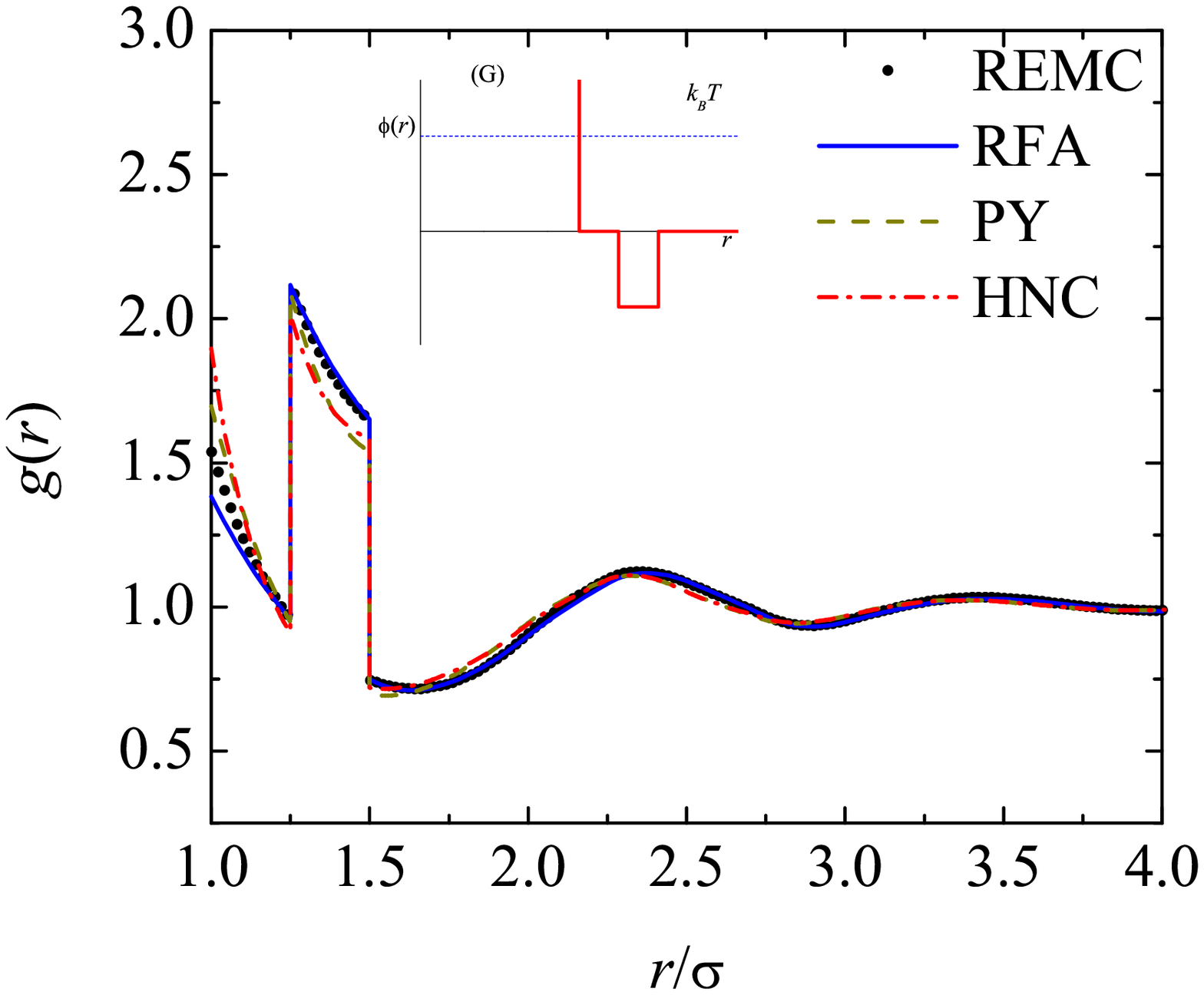}
\caption{ Comparison of the different theoretical approaches to compute the RDF of the system corresponding to case G ($\lambda_1=1.25$, $\la_2=1.5$, $\epsilon_1=0$, $\epsilon_2=-\epsilon$) at $\rho^*=0.5$ and $T^*=1.26193$ with simulation results. Solid line: RFA; dashed line: HNC; dotted-dashed line: PY; symbols: REMC data. The inset shows the interaction potential and the temperature value.}
\label{fig9}
\end{figure}

\begin{figure}
\includegraphics[width=8cm]{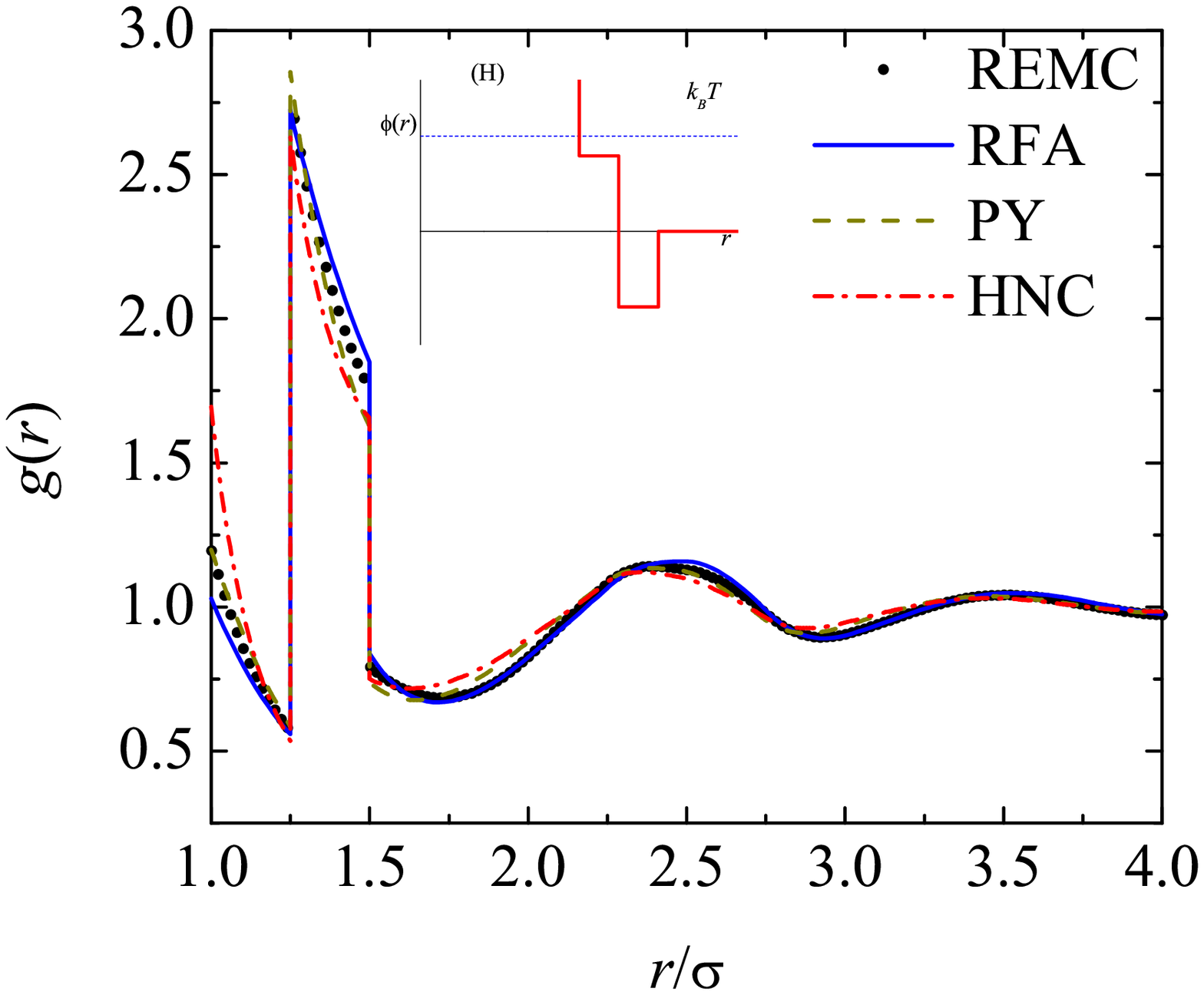}
\caption{ Comparison of the different theoretical approaches to compute the RDF of the system corresponding to case H ($\lambda_1=1.25$, $\la_2=1.5$, $\epsilon_1=\epsilon$, $\epsilon_2=-\epsilon$) at $\rho^*=0.5$ and $T^*=1.26193$ with simulation results. Solid line: RFA; dashed line: HNC; dotted-dashed line: PY; symbols: REMC data. The inset shows the interaction potential and the temperature value.}
\label{fig10}
\end{figure}

As an illustration of the results of our calculations, and in accordance with the simulation experiments mentioned above,
we fix the reduced density to be $\rho^*=0.5$ and consider the {lowest} reduced temperature $T^*=1.26193$ {(corresponding to $n=M=12$)} in all cases, although, {as said before,} we obtained simulation data for the whole temperature range $1.26193\le T^* \le 2$.
The results {for the RDF} are displayed in Figs.\ \ref{fig3}--\ref{fig10} for cases A--H, respectively. {In order to assess the influence of $\epsilon_1$ and $\epsilon_2$ on $g(r)$ at fixed (reduced) temperature and density, common horizontal and vertical scales have been chosen in Figs.\ \ref{fig3}--\ref{fig10}. In the case of the purely repulsive system A, there exists a local accumulation of particles at the external edge ($\lambda_i^+$) of each repulsive step, followed by a local depletion at the internal edge ($\lambda_i^-$).  For systems B--D, where the potential is repulsive at $r/\sigma=\la_2$ but attractive at $r/\sigma=\la_1$, the population of particles (as seen by a reference particle at the origin) is depleted in the  region $\la_1<r/\sigma<\la_2$ and increases when going from $\lambda_1^+$ to $\lambda_1^-$, as expected. These effects are enhanced as the depth of the inner well increases.
In the case of system E the potential outside the hard core is purely attractive, what is reflected in an increase of $g(r)$  from the external edge $\lambda_i^+$ to the internal edge $\lambda_i^-$. Finally, systems F--H are the counterparts of systems B--D. Now the attraction at $r/\sigma=\la_2$  produces an increase of particles in the region $\la_1<r/\sigma<\la_2$, this effect being enhanced as the inner barrier becomes more repulsive.}

{Let us comment now on the theoretical predictions.} It follows that, as already pointed out in Ref.\ \onlinecite{SYH12}, the RFA approach certainly outperforms the PY approximation in all the cases. This is noteworthy because, while  the RFA reduces to the PY solution for hard spheres,\cite{SYH12} it is much simpler than the PY integral equation theory for two-step potentials. As for the HNC integral equation theory, it presents the best agreement in the region $1<r/\sigma<\la_1$ in the cases A--C, i.e., when $\epsilon_1\geq 0$ and $\epsilon_2>0$. Even in those cases, however, the RFA is as accurate as or more accurate than the HNC theory in the region $\la_1<r/\sigma<\la_2$. For larger distances the RFA and HNC predictions are almost indistinguishable. In the rest of the cases (D--H), the PY and HNC curves are generally very similar, the best global performance being obtained with the RFA.

{An additional advantage of the RFA over the numerical solutions of integral equations is that, since the $s$-dependence of the Laplace transform $G(s)$ is fully explicit, the correlation length of the system can be straightforwardly obtained.
This relies upon the search for the pole (or conjugate pair of poles) $s=-\kappa\pm i\omega$ of $G(s)-s^{-2}$  with the negative real part $-\kappa$ closest to the origin. As a consequence, the asymptotic behavior of $h(r)$ is given by}
\beq
{h(r)\sim  \frac{e^{-\kappa r}}{r} \cos(\omega r+\phi),}
\label{corr}
\eeq
{where $\kappa$ is the inverse correlation length and $2\pi/\omega$ is the wavelength of the oscillations.
At the common thermodynamic state $\rho^*=0.5$ and $T^*=1.26193$, we have found $(\kappa,\omega)=(1.503,5.128)$, $(1.827, 4.424)$, $(1.704,7.116)$, $(1.378, 6.990)$, $(1.327, 6.225)$, $(1.059, 5.856)$, $(0.955, 5.738)$, and $(0.754, 5.632)$ for systems A--H, respectively. Therefore, the smallest correlation length $\kappa^{-1}\simeq 0.55$ corresponds to case B and it monotonically increases from system B to system H, where in the latter case one has $\kappa^{-1}\simeq 1.33$. Case A, with $\kappa^{-1}\simeq 0.67$, lies in between cases C and D. The wavelength has a less systematic behavior, ranging from $2\pi/\omega\simeq 0.88$ (case C) to  $2\pi/\omega\simeq 1.42$ (case B).
Interestingly enough, although Eq.\ \eqref{corr} applies to the asymptotic regime $r\to\infty$ only, the increase of the correlation length when going from B to H agrees with what is observed in Figs.\ \ref{fig4}--\ref{fig10}, where the distance beyond which $|g(r)-1|\leq 0.03$ turns out to be $2.1$, $2.45$, $2.49$, $2.54$, $2.76$, $3.01$, $3.08$, and $3.71$ for systems A--H, respectively.}

\begin{figure}
\includegraphics[width=8cm]{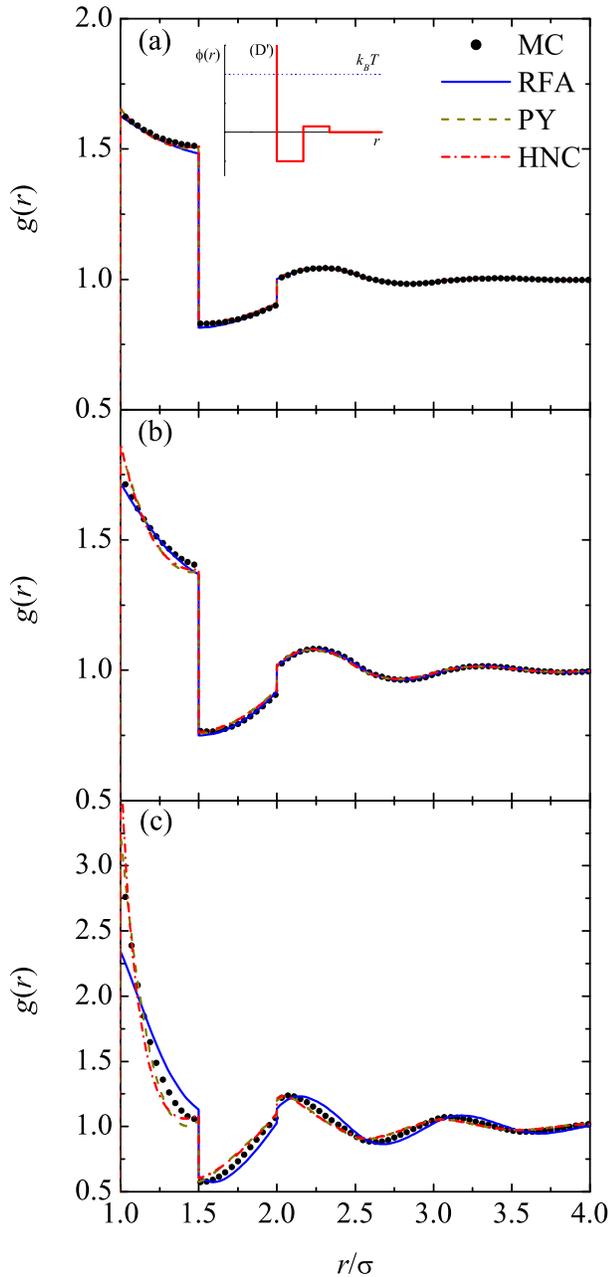}
\caption{ Comparison of the different theoretical approaches to compute the RDF of the system corresponding to case D' ($\lambda_1=1.5$, $\la_2=2$, $\epsilon_1=-\epsilon$, $\epsilon_2=\epsilon/5$) at $T^*=2$ and (a) $\rho^*=0.2$, (b) $\rho^*=0.4$, and (c) $\rho^*=0.75$ with simulation results. Solid line: RFA; dashed line: HNC; dotted-dashed line: PY; symbols: MC data.\protect\cite{HTS11} The inset shows the interaction potential and the temperature value.}
\label{fig11}
\end{figure}

\begin{figure}
\includegraphics[width=8cm]{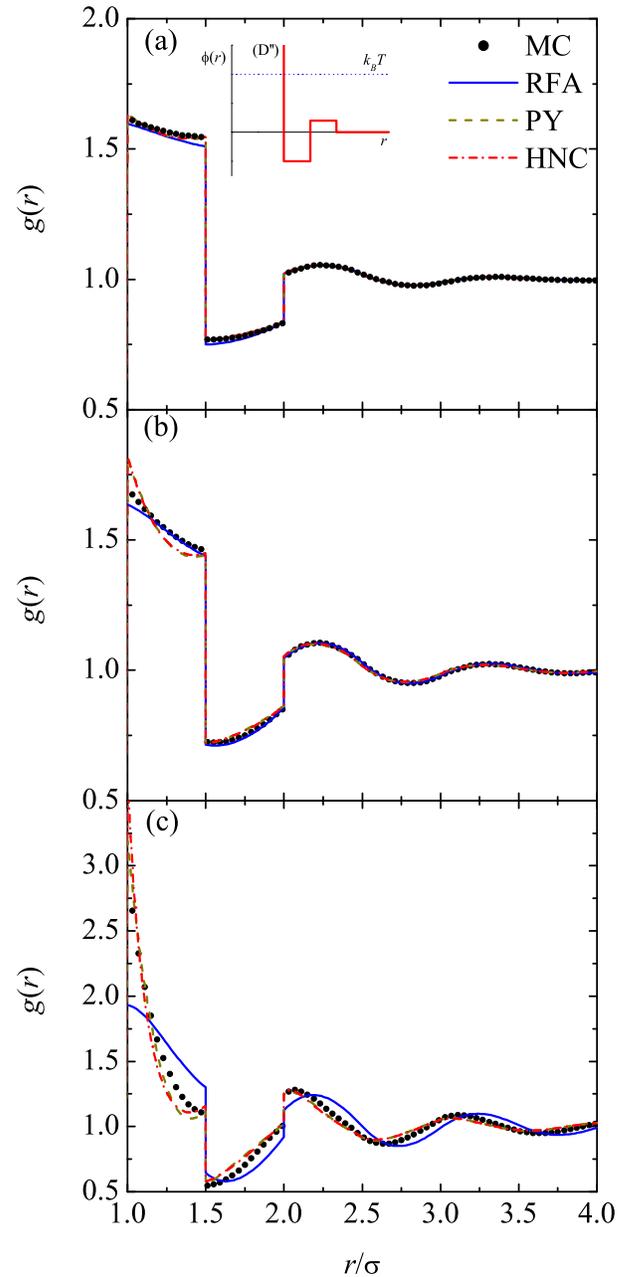}
\caption{ Comparison of the different theoretical approaches to compute the RDF of the system corresponding to case D'' ($\lambda_1=1.5$, $\la_2=2$, $\epsilon_1=-\epsilon$, $\epsilon_2=2\epsilon/5$) at $T^*=2$ and (a) $\rho^*=0.2$, (b) $\rho^*=0.4$, and (c) $\rho^*=0.75$ with simulation results. Solid line: RFA; dashed line: HNC; dotted-dashed line: PY; symbols: MC data.\protect\cite{HTS11} The inset shows the interaction potential and the temperature value.}
\label{fig12}
\end{figure}

As a final illustration, we present in Figs.\ \ref{fig11} and \ref{fig12} a comparison of the results we have obtained with the different theoretical approaches with those of MC simulations\cite{HTS11} of two systems with an intermolecular potential of the type of case D (square well + square barrier), but this time having the following values of the parameters: $\lambda_1=1.5$, $\lambda_2=2$, $\epsilon_1=-\epsilon$, $\epsilon_2=\epsilon/5$ (system D') and $\lambda_1=1.5$, $\lambda_2=2$, $\epsilon_1=-\epsilon$, $\epsilon_2=2\epsilon/5$ (system D''), respectively. For those systems we have considered a  fixed temperature $T^*=2$ and the three reduced densities $\rho^*=0.2$, $\rho^*=0.4$, and $\rho^*=0.75$.\cite{note_13_04}
At the lowest density ($\rho^*=0.2$) the three theoretical approaches provide excellent results, with a slight superiority of PY and HNC. {This phenomenon is a reflection of the fact that, as said above, the PY and HNC approximations are  exact to order $\rho$ while the RFA is not.} At the intermediate density ($\rho^*=0.4$), however, the RFA beats the PY and HNC results (which are practically indistinguishable from each other). Finally, at the highest density ($\rho^*=0.75$) the RFA becomes clearly worse than the PY and HNC predictions (which are again hardly distinguishable), especially in the case of the higher barrier (case D''). {The shortcomings of the RFA as the density and the widths of the potential sections  increase have been reported for the square-well case in Ref.\ \onlinecite{YS94}.}

\section{Concluding remarks}
\label{sec4}

In this paper we have presented a rather systematic study of the RDF of fluids whose molecules interact via a potential with a hard core plus two piece-wise constant sections of the same width and different heights (either wells or shoulders), which include the RFA approach, our REMC numerical experiments, and the numerical solution of the PY and HNC integral equations. We have considered eight representative classes of systems (see Fig.\ \ref{fig2}) which cover all the possible topologies of hard core plus two-step potentials. They include a purely repulsive potential (system A), potentials with an outer repulsive barrier and an inner attractive well (systems B--D), a purely attractive part outside the hard core (system E), and potentials with an outer attractive well and an inner repulsive barrier (systems F--H). {Four of these systems (A and F--H) belong to the class of core-softened potentials.}

As Figs.\ \ref{fig3}--\ref{fig10} show, it is fair to state that for the fixed number density $\rho^*=0.5$ the agreement between the results of our RFA formulation and those of the REMC experiments is very satisfactory in all instances. This is specially rewarding in view of the fact that the reduced temperature we have chosen to illustrate our findings ($T^*=1.26193$) is rather low and represents a stringent test of our theory.

A specially relevant point is the capability of the RFA to provide the correlation length from the  pole of the Laplace transform $G(s)$ with the negative real part closest to the origin. The fact that, at the given state point,  the potential H has the largest correlation length (larger than $1.3$) is consistent with the results of Ref.\ \onlinecite{VF10} for a similar (continuous) potential, where  the (anomalous) thermodynamic behavior of the system was shown to be determined by the contribution to the RDF coming from up to the fourth coordination shell.

We have complemented our study with the comparison of the results we get and the MC data of Ref.\ \onlinecite{HTS11} for two cases (labeled here as D' and D'') in which the potential is of the form of the one of case D but with a wider width, a lower outer barrier, and a fixed reduced temperature $T^*=2$. Again, as observed from Figs.\ \ref{fig11} and \ref{fig12}, the performance of the RFA approach is rather satisfactory, except at the highest density $\rho^*=0.75$. The tendency of the RFA to fail at high densities and wide potentials widths was already documented in the case of the pure square-well potential (with $\epsilon_2=0$).\cite{YS94}

Concerning the comparison between our present approach and the usual integral equation approach in the theory of liquids, we have seen that the RFA is rather simple, requires much less numerical labor (in these cases only the solution of two coupled transcendental equations), captures correctly all the oscillations of the RDF, and is reasonably accurate. In fact, it is always superior to the PY equation [except for the cases of Figs.\ \ref{fig11}(c) and \ref{fig12}(c)] and in most of the instances it is of comparable accuracy or better than the HNC equation. Hence, this constitutes further evidence of the usefulness of the RFA methodology that we have used for the computation of the structural properties of {different} hard-core fluids.\cite{HYS08}

{All the calculations that we have presented have been made at temperatures higher than those of the possible vapor-liquid and liquid-liquid phase transitions in these systems.\cite{FMSBS01,SBFMS04,MFSBS05,FRT06} An interesting problem is the description of the structure of such systems in the vicinity of these transitions. The availability of the analytical results for the structural properties in Laplace and Fourier spaces as obtained from the RFA approach allows one to tackle this problem using the same procedure that was applied in the case of the critical point of the square-well fluid.\cite{AS01} Also of interest in connection not only with these phase transitions but in a more general perspective  is the study of the thermodynamic properties of the systems, in particular the equation of state, through the virial, energy, and compressibility routes. Again the RFA approach permits such a determination. Work along these two lines is in progress and will be reported elsewhere.
}

As a final point, we want to stress that the results of this paper, together with the earlier ones,\cite{SYH12} encourage us to consider the problem in which the number of steps in the potential is much greater, leading in the limit $n \to \infty$ to the very interesting case of the structure in a fluid whose molecules interact  {via, for instance,} a Jagla potential. We plan to undertake such a task in the future.

\acknowledgments

We want to thank R. Casta{\~n}eda-Priego  and S. P. Hlusak for kindly supplying us with all the simulation data of Refs.\ \onlinecite{GCC07} and \onlinecite{HTS11}, respectively. Two of us (A.S. and S.B.Y) acknowledge the financial support of the Spanish Government through Grant No.\ FIS2010-16587 and  the Junta de Extremadura (Spain) through Grant No.\ GR10158 (partially financed by FEDER funds). P.O. thanks the Molecular Engineering Program of IMP.



\begin{thebibliography}{39}
\bibitem[{not({\natexlab{a}})}]{note_13_04_2}
\bibinfo{note}{{More specifically, we have used the command FindRoot of
  Mathematica.}}
\bibitem[{not({\natexlab{b}})}]{note_13_07}
\bibinfo{note}{The optimal values of the mixing parameter $\alpha$ are
  generally slightly below the upper bound values for convergence. Typical
  values for the systems A--H analyzed in Sec.\ \protect\ref{sec3} are $0.2$ (A
  and B), $0.3$ (C and H), $0.4$ (G), $0.5$ (F), and $0.6$ (D and E). As for
  the systems D' and D'', the optimal values of $\alpha$ are typically $0.2$
  ($\rho^*=0.75$), $0.7$ ($\rho^*=0.4$), and $0.9$ ($\rho^*=0.2$).}
\bibitem[{not({\natexlab{c}})}]{note_13_04}
\bibinfo{note}{{Completely similar systems were considered in Ref.\
  \protect\onlinecite{GCC07} but for different reduced densities. For brevity
  and because they are more recent, we decided to present here only the
  comparison with the results of Ref.\ \protect\onlinecite{HTS11}. However, we
  have also performed a comparison between our results and the former MC data
  and this comparison is available upon request.}}
\end{thebibliography}


\end{document}